\documentclass[aps, prd,11pt,reprint,twocolumn]{revtex4-1}
\usepackage[english]{babel}
\usepackage[utf8]{inputenc}
\usepackage{amsmath}
\usepackage{amssymb}
\usepackage{amsthm}
\usepackage{mathtools}
\usepackage{dsfont}
\usepackage{graphicx}
\usepackage{soul}
\usepackage{xcolor}

\newcommand{\ket}[1]{\left|{#1}\right\rangle}
\newcommand{\bra}[1]{\left\langle{#1}\right|}

\renewcommand{\Re}{\operatorname{Re}}

\newtheorem{theorem}{Theorem}
\newtheorem{postulate}{Postulate}

\newtheorem{corollary}{Corollary}

\begin{document}
\title{Relativistic time dilation from a quantum mechanism}
\author{Esteban Mart\'inez Vargas}
\email{estebanmv@protonmail.com}
\affiliation{Calle centella 3, Smza 18 Mza 3, 77505 Cancún México}

\begin{abstract}
    One of the concepts of Relativity theory that challenges conventional intuition the most
    is time dilation and length contraction. 
    Usual approaches for describing relativistic effects in quantum systems merely postulate 
    the consequences of these effects as physical constraints.
    Here, we propose to rebuild Special Relativity from quantum mechanical considerations.
    This is done by dropping one of its fundamental postulates: the
    universality of the speed of light. Lorentz transformations are obtained by
    a quantum mechanism. We use the fact that quantum states depend on the
    Galilean reference frame where they are defined. In other words, quantum
    states outside an observer's Galilean reference frame are distorted.
    Then, we show in a theorem the existence of time-dependent observables 
    that are sensible to this distortion in such a way that their 
    expectation value is a Lorentz-covariant function of time.
    We then postulate this mechanism as the source of the phenomena of Special
    Relativity.
    As a corollary of the main theorem, we show the existence of a Lorentz covariant momentum and mass operators
    which yield the relativistic momentum and mass.
    In this theory, the fundamental limit of the speed of light imposes a 
    transparency condition for faster-than-light
    particles: they are allowed but they are not observable. 
    The transparency effect could explain dark matter in a more general
    theory following this quantum formalism.
\end{abstract}
\maketitle
\section{Introduction}

One of the deepest questions in contemporary physics is that of quantum gravity. 
A universal phenomenon such as gravitation should have a description in terms 
of the most
fundamental theory at hand: quantum theory \cite{sakurai1985modern}.
Despite big efforts, it has remained a mystery how to build a successful quantum 
theory of gravitation~\cite{rovelli_2004}. 

As it is universally known, the most accurate description of gravity is given by
Einstein's General Theory of Relativity, of which Special Relativity is a particular case~\cite{wald2010general}.
Although there have been successful theories that have merged the phenomenology of Special
Relativity with quantum mechanics as Quantum Field Theory and Dirac's equation~\cite{itzykson2012quantum,greiner1996field}, it has remained 
elusive how to do this in the general theory. Much of it has to do with the fact
of how such theories treat the concept of spacetime, which is central to General Relativity. 
For example, one of the challenges resides in the fact that quantum phenomena are taken with an absolute
causal structure whereas General Relativity does not have one 
\cite{TowardsQuantumHardy2007,BellsTheoremZych2019}.

This could point toward the abandonment of the use of absolute time frames
to describe phenomena in quantum systems \cite{EvolutionWithoPage1983,QuantumComputaChirib2013,ANoGoTheoremCosta2022,BellsTheoremZych2019}. 
If absolute time is not necessary it is however a powerful concept to describe quantum 
phenomena. In any case, this asks for a build-up of a theory of spacetime for quantum systems.

The usual \emph{ontology} of spacetime is some kind of field that permeates
the universe which interacts with matter slowing down time and expanding space 
accordingly. This is the viewpoint of approaches 
that originate from particle physics like Loop Quantum gravity~\cite{rovelli_2004} and 
the AdS/CFT correspondence~\cite{Maldac1999}. 
Recently, there has been an interest in using the tools from 
low energy quantum theory and quantum information theory 
to study relativistic effects~\cite{QuantumInformaPeres2004,QuantumFormulaZych2018,AliceFallsIntFuente2005}. 
However, these low-energy studies are from the
same ontological point of view. In all these approaches Relativity theory has a superior 
status in some sense: it has some symmetry properties that quantum systems 
are constrained to fulfill. 

\begin{figure}
    \includegraphics[width=0.49\textwidth]{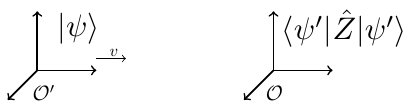}
    \caption{State $\ket{\psi}$ in reference frame $\mathcal{O}^\prime$
        turns into state $\ket{\psi^\prime}$ in reference frame $\mathcal{O}$.}
    \label{fig:statesAndRFs}
\end{figure}

We propose to build a theory of spacetime from purely quantum considerations.
The canonical approaches to recover relativistic covariance are to impose it
in the solutions of the Dirac equation and on the field operators in Quantum
Field Theory~\cite{itzykson2012quantum}. 
We propose a somewhat similar approach to Quantum Field Theory: we impose
a relativistic covariance in the operators.
We have a velocity-dependent state $\ket{\psi}$ from a reference frame 
$\mathcal{O}^\prime$.
Applying a transformation we obtain $\ket{\psi^\prime}$
in the reference frame $\mathcal{O}$. An observable is taken in 
reference frame $\mathcal{O}$, this is shown in Fig. (\ref{fig:statesAndRFs}).
We show that covariant observables $\hat{Z}$ exist. In other words, we
\emph{modify the operators} so that they become Lorentz covariant. 

The main difference with the usual Quantum Field Theory is that the transformation from $\ket{\psi}$ to
$\ket{\psi^\prime}$ is a \emph{Galilean transformation}. We recover
relativistic covariance from the observables $\hat{Z}$. 
%
We rebuild Special Relativity with quantum principles.
As it is known, Lorentz transformations of Special Relativity 
are deduced from two fundamental 
postulates \cite{schutz_2009}: the Galilean principle 
of relativity and the universality of the speed of light. Here we
do not use the universality of the speed of light to derive Lorentz
transformations, we pose instead the following postulate which has a quantum
origin.
\begin{postulate}
    Matter in a reference frame distorts matter from
    other reference frames in such a way that information acquisition follows
    Lorentz transformations.
    \label{pst:QLrntzPost}
\end{postulate}

This all comes from the fact that quantum states depend on the Galilean reference
frame they are defined in \cite{QuantumMechaniGiacom2019}. Thus, seen from a reference
frame all quantum states from other reference frames are distorted. Here, we only show
that there are observables that are sensible to this distortion and whose
value grows in time according to Lorentz transformations. We thus state that
these observables are present in all matter interactions.

%
The Galilean transformation for quantum states was previously introduced 
in~\cite{OnUnitaryRayBargma1954}
and extended in~\cite{SomeRemarksOnGreenb1979,QuantumMechaniGiacom2019}.
We show that there exist observables that grow linearly in time which, in 
the other reference frame transform in a way that makes time dilate. Specifically, the 
expectation value would grow linearly but with a flatter slope that accounts for Lorentz 
time dilation.

To understand the proposed mechanism it is useful to think in terms 
of quantum machines that acquire information.
Quantum devices have been
used to create regular motion that produces ticks as we have seen in atomic clocks \cite{HamiltonianEngAeppli2022}
and also in time crystals \cite{QuantumTimeCrWilcze2012}.
The analysis of these devices yields deep insights into the relationship of
time and matter \cite{AutonomousQuanErker2017}.
However, the second part of the clock, the record is as important. 
The problem is that, if no record is taken, then what does regularity mean in the 
first place? We have to remember that a motion has already occurred so that
we can say that it has happened again.
This is a crucial point here, as we will not describe a quantum clock but a \emph{quantum time register}.
This means the mechanism that indicates time passage in accumulation, similar to an 
hourglass. The concept of time register will be formalized in the section \ref{sec:CTR}, it is based
on the theory of Sequential Analysis~\cite{wald1973sequential}. Afterward, we extend this
concept to Quantum Mechanics in section \ref{sec:QTR}. 

We then study in section \ref{sec:GalTrans} the Galilean transformations
amongst reference frames. We observe that a transformed state $\ket{\psi^\prime}$
has the information about the relative velocity of the frame $\mathcal{O}^\prime$.
We also observe that the momentum $\hat{p}$ and position $\hat{x}$ operators concerning the transformed states reproduce exactly the classical momentum and position. 
This suggests a relativistic momentum operator $\hat{p}^{Rel}$ and a relativistic mass operator $\hat{\mu}^{Rel}$,
which are defined in section \ref{sec:RelCovar}. In that section, we also
define the Lorentz covariant QTRs.

We study observables whose expectation value grows linearly in time, denoted as 
$\hat{Z}(t)$. 
We prove in section \ref{sec:LorCovQTR} that there can exist 
observables whose expectation value grows linearly and have a slope according 
to Lorentz transformations, this is, that grows as $\epsilon t/\gamma$ for a constant
$\epsilon$, and $\gamma$ is the Lorentz factor.
This is theorem \ref{thm:linealgrowth}. The
observable $\hat{Z}$ does not depend on velocity but the states $\ket{\psi^\prime}$ do
depend on it. We have that $\bra{\psi^\prime}\hat{Z}\ket{\psi^\prime}=\epsilon t/\gamma$.
Therefore $t=\gamma t_0$.

We then pose in section \ref{sec:AQOrigin} that there exists an 
hourglass-like mechanism in matter which yields spacetime.
This is done by showing that for all Lorentz covariant holomorphic 
functions of time $g(t/\gamma)$ there is an observable $\hat{Z}_g$ such that
$\bra{\psi^\prime}\hat{Z}_g\ket{\psi^\prime}=g(t/\gamma)$.
This is theorem \ref{thm:holofunc}.
This is an alternative derivation of Special Relativity.
We summarize this theory in postulate \ref{pst:QTRpost} which is an alternative form
of postulate \ref{pst:QLrntzPost}.

In section \ref{sec:Cors} we prove two corollaries of theorem \ref{thm:holofunc}.
Corollaries \ref{cor:RelMom} and \ref{cor:RelMass} show the existence of the 
relativistic momentum $\hat{p}^{Rel}$ and mass operators $\hat{\mu}^{Rel}$ 
respectively, defined in section \ref{sec:RelCovar}.

Exploring this line of thought we realize that the constraint of the speed of
light as limiting speed is needed. We note in section \ref{sec:TransAndTwin}
that in this theory such a limit would translate into a transparency
condition. This means, in this theory, faster-than-light particles are allowed, however, they become
invisible to certain observers.

Section \ref{sec:disc} is the discussion where we observe that 
this theory is not against a field in General
Relativity: it merely states that an interaction of the field and the observables
$\hat{Z}_g$ has to exist. Furthermore, a Generalized theory could help explain
dark matter as a consequence of the transparency phenomenon.
\section{Classical time register}
\label{sec:CTR}
\begin{figure}
    \includegraphics[width=0.5\textwidth]{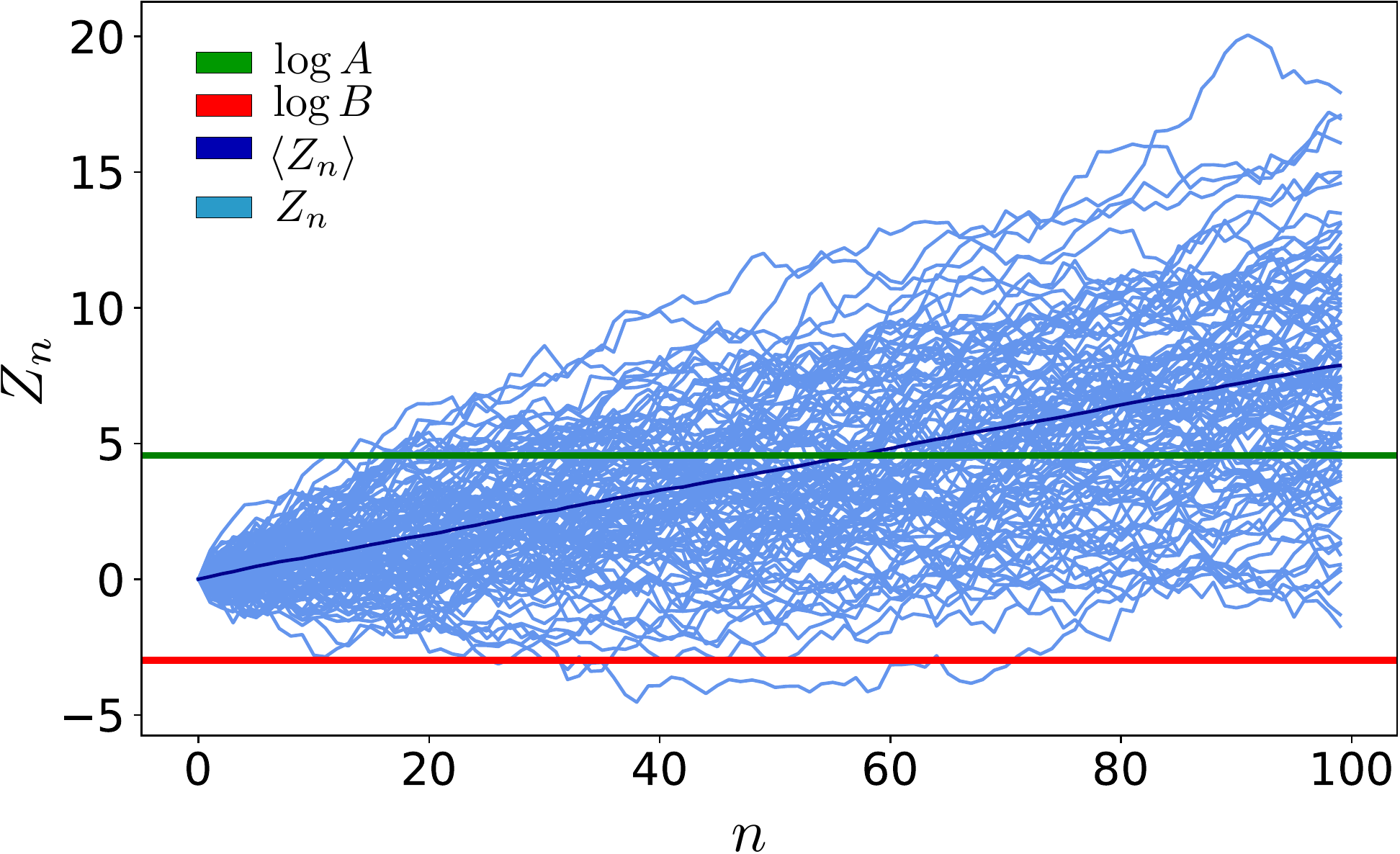}
    \caption{Martingale example with 1000 numerical experiments. In the horizontal 
        axis, we have the number of iterations from the Wiener process. The vertical axis
        contains the total value of the sum of the random walks. We see that the
        process is biased upwards. The deep blue line is the numerical mean of the 
        1000 random walks. We see that the mean grows linearly in time.}
    \label{fig:mrtngl}
\end{figure}

The Sequential Probability Ratio Test (SPRT) is the optimal protocol in terms of the resources needed for 
the acquisition of information to decide between two hypotheses and is the central protocol
studied by Wald~\cite{wald1973sequential}. Specifically, given an error threshold, it is the protocol
that minimizes the needed resources to make a decision. Surprisingly, the central object
in this protocol is not deterministic at all: it is a random walk \cite{van1992stochastic,mitzenmacher_upfal_2005,CoverThomasElements2006}. 
In the important Gaussian case, it is a Wiener process \cite{jacobs_2010,wiseman_milburn_2009,SequentialAnalMartin2022} that
will stop until a threshold is reached. However, this process happens \emph{in time}, which
means, it is correlated with time.
Therefore, we can correlate Sequential Analysis 
not as a clock but as a \emph{time register}. This means, that if we assume that the samples
detected are exactly correlated with the ticks of a perfect clock, we can use the random walk
as evidence of the passage of time.
In Fig. (\ref{fig:mrtngl}) we have 1000 samples of a Wiener process with $N=100$ that
is biased. The information about one hypothesis or another comes from this bias. The
average of the random walkers is a line with a specific slope that depends on the
sampling distribution.
We can write the evolution of $Z$ as a Wiener process so it 
obeys a dynamical equation,
\begin{equation}
    Z(t+dt)-Z(t)=\sqrt{\delta^2dt}N_t^{t+dt}(\theta,\frac{1}{2}).
\end{equation}
Where $N_t^{t+dt}(\theta,\frac{1}{2})$ represents a Gaussian with mean $\theta$ and
variance $1/2$.
Observe that this is an instance of a stochastic differential equation that obeys
different rules from usual calculus~\cite{jacobs_2010}.
In the SPRT, when the random walker crosses a certain threshold, a decision 
with certain error bounds can be made.
\section{Quantum time register}
\label{sec:QTR}
The previous paragraph suggests that for the acquisition of information one should use a 
random walker. However, it should be a biased one, so that the general trend evidences
the acquisition of information. Moreover, from Fig. (\ref{fig:mrtngl}) we see that the
trend is that the mean value of the Martingale grows linearly in time, $\epsilon t$ for
some constant $\epsilon$. Can we extend the previous concept of time registers for 
quantum systems? The quantum analog of $Z$ would be an observable that evolves with $t$.

\begin{figure}
    \includegraphics[width=0.5\textwidth]{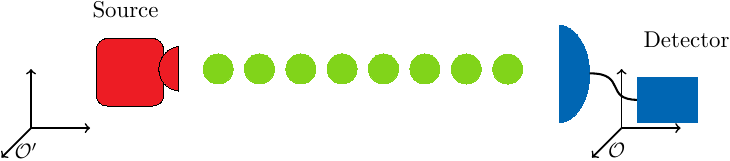}
    \caption{IID source produces equally spaced states $\ket{\psi_0}$. The source
        and the detector can be in different reference frames.}
    \label{fig:iidsrc}
\end{figure}

In contrast with time crystals the value of the observable should always
grow. Suppose that we have a perfectly regular clock and that at each tick it
yields a quantum state $\ket{\psi_0}$. This state travels and a measurement is performed 
on the observable $\hat{Z}(t)$ as depicted in Fig. (\ref{fig:iidsrc}). As quantum measuring is a stochastic process, we would
have something similar to Fig. (\ref{fig:mrtngl}). The bias should be such that it
has to be proportional to the passage of time. Following the analogy with the martingales of Fig. (\ref{fig:mrtngl})
we define a Quantum Time Register (QTR) as any observable $\hat{Z}(t)$ such that the expectation
value concerning $\ket{\psi_0}$ is proportional to time, considering the difference with
an initial $t_0$ i.e.
\begin{equation}
    \bra{\psi_0}\left(\hat{Z}(t)-\hat{Z}(t_0)\right)\ket{\psi_0}=\epsilon\Delta t.
    \label{eq:Ztime}
\end{equation}

\section{Galilean transformations of quantum states}
\label{sec:GalTrans}
From the formal studies of representation theory by Bargmann \cite{OnUnitaryRayBargma1954}
it was observed that
quantum states are altered when considered in different Galilean frames of reference.
Afterward, Greenberger \cite{SomeRemarksOnGreenb1979} extended these transformations and proposed a modified 
Schrödinger equation that includes a term which corresponds to a possible gravitational
potential. Greenberger also provided physical intuition to a mass superselection rule by
Bargmann. Then Giacomini et. al. \cite{QuantumMechaniGiacom2019} studied general transformations between
reference frames. 

A relevant aspect to note is that all these are transformations of states. The
operators are left intact. Explicitly, to change a quantum particle description 
$\ket{\psi}$ in reference frame $\mathcal{O}$ to the state $\ket{\psi^\prime}$
described in the reference frame $\mathcal{O}^\prime$ which is at constant velocity
$v$ with respect to $\mathcal{O}$ we apply the following transformation, 
\begin{equation}
    \ket{\psi^\prime}=e^{\frac{i}{\hbar}v\hat{G}}\ket{\psi}.
\end{equation}
As the velocity $v$ depends on how one chooses the direction of the reference frame
we will conveniently choose to make the change $v\rightarrow-v$.
The generator of the galilean boost is denoted $\hat{G}:=\hat{p}t-m\hat{x}$ with $m$
the mass of the particle $\ket{\psi}$. 
Expressing the state $\ket{\psi}$ in the basis of the harmonic oscillator is very
convenient. Observe, that, for the harmonic oscillator the operators of momentum
and position are defined as
\begin{equation}
    \hat{x} = \sqrt{\frac{\hbar}{2m\omega}}(\hat{a}^\dagger+\hat{a})\quad\quad\hat{p} = i\sqrt{\frac{\hbar m\omega}{2}}(\hat{a}^\dagger-\hat{a}),
\end{equation}
for an angular frequency $\omega$ and $\hat{a}$ and $\hat{a}^\dagger$ the creation and
annihilation operators respectively. The transformation between Galilean reference
frames become
\begin{equation}
    e^{-\frac{i}{\hbar}v(t\hat{p}-m\hat{x})} = e^{\alpha\hat{a}^\dagger-\alpha^*\hat{a}},
\end{equation}
for 
\begin{equation}
    \alpha\equiv v\sqrt{\frac{m}{2\hbar}}(t\sqrt{\omega}+\frac{i}{\sqrt{\omega}}).
    \label{eq:alphadef}
\end{equation}
Observe that $e^{\alpha\hat{a}^\dagger-\alpha^*\hat{a}}$ is a displacement operator
\cite{scully_zubairy_1997}. It will be very convenient to choose $\ket{\psi}=\ket{0}$,
because applying a Galilean transformation yields a coherent state,
\begin{equation}
    \ket{\alpha} = e^{\alpha\hat{a}^\dagger-\alpha^*\hat{a}}\ket{0}.
\end{equation}
This will be, of course, a harmonic oscillator with mass.

The state $\ket{\alpha}$ has the information about the velocity of the
reference frame $\mathcal{O}^\prime$. Observe that the expected value of the 
momentum with respect to $\ket{\alpha}$ is
\begin{equation}
    \langle\alpha|\hat{p}|\alpha\rangle =  i\sqrt{\frac{\hbar m\omega}{2}}(\alpha^*-\alpha) = vm,
    \label{eq:expecmomentum}
\end{equation}
which coincides exactly with the classical definition of momentum. 
Observe that for the position operator we have
\begin{equation}
    \langle\alpha|\hat{x}|\alpha\rangle = \sqrt{\frac{\hbar}{2m\omega}}(\alpha^*+\alpha) = vt,
\end{equation}
which is exactly the classical definition of the position of the origin of reference frame $\mathcal{O}^\prime$
with respect to the origin of $\mathcal{O}$.
\section{Relativistic covariance}
\label{sec:RelCovar}
As we have noticed before the momentum operator is never modified
in relativistic quantum mechanics. Here we \emph{modify the observables} so that
the relativistic effects come from them. Following the result from the momentum
operator in Eq. (\ref{eq:expecmomentum}) we introduce 
a relativistic momentum operator $\hat{p}^{Rel}$ such that
\begin{equation}
    \langle\alpha|\hat{p}^{Rel}|\alpha\rangle = \frac{vm}{\sqrt{1-\frac{v^2}{c^2}}~}.
    \label{eq:RelMom}
\end{equation}
In a similar manner we introduce a covariant mass operator $\hat{\mu}_r$ such that
\begin{equation}
    \langle\alpha|\hat{\mu}^{Rel}|\alpha\rangle = \frac{m}{\sqrt{1-\frac{v^2}{c^2}}~}.
    \label{eq:RelMass}
\end{equation}
We will show the existence of $\hat{p}^{Rel}$ and $\hat{\mu}_r$ in the following 
sections. 

Following these lines of thought, if relativistic effects are caused by the observables
then we would have to propose how time dilation occurs in this scheme. This
involves the study of QTRs.
%

\begin{figure}
    \includegraphics[width=0.5\textwidth]{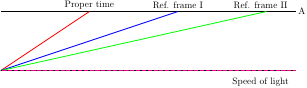}
    \caption{Different reference frames imply a different growth in time.}
    \label{fig:avsgrow}
\end{figure}

We want a QTR that is consistent with relativistic time dilation. 
Such observable $\hat{Z}$ would have the expectation value
\begin{equation}
    \bra{\psi^\prime}\left(\hat{Z}(t)-\hat{Z}(t_0)\right)\ket{\psi^\prime}=\frac{\epsilon}{\gamma}\Delta t,
    \label{eq:ZtimeLntz}
\end{equation}
where 
\begin{equation}
    \gamma := \frac{1}{\sqrt{1-\frac{v^2}{c^2}}~}
\end{equation}
is the Lorentz factor and $\Delta t=t-t_0$. In what follows we will take units where $c=1$. 
We are searching for an observable, which implies a mechanism in the 
interaction that gives rise to time dilation
as illustrated in Fig. (\ref{fig:avsgrow}). An operator with an expectation value 
that grows
the proper time frame would behave as in the red line until it reaches the value $A$. 
In a different reference frame, the growth would take longer, as with the blue and green lines. If the speed is
too high the expectation value would never reach the value $A$.

In other words, we want to see if an observable with the property of Eq. ($\ref{eq:ZtimeLntz}$) at
a given time can differentiate amongst reference frames. The observable will be sensible to 
the states coming from different reference frames, which depend on the relative velocity.
The expectation value will thus be a function of velocity $\langle\hat{Z}\rangle=f(v)$ at a given time,
therefore marking the difference between reference frames. However, $\hat{Z}$
cannot depend on $v$, the $v$ dependence comes from the state $\ket{\psi^\prime}$.
\section{Existence of Lorentz-covariant QTRs}
\label{sec:LorCovQTR}
%
We will use the coherent states previously introduced, i.e. $\ket{\psi}^\prime=\ket{\alpha}$.
The question is now if there exists an observable $\hat{Z}$ that grows linearly in time
according to the Lorentz transformations as in Eq. (\ref{eq:ZtimeLntz}).
We show the existence of such observables in the following theorem.

\begin{figure}
    \includegraphics[width=0.4\textwidth]{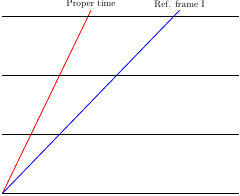}
\caption{Several clock ticks in different reference frames.}
\label{fig:clockticks}
\end{figure}

\begin{theorem}
    \label{thm:linealgrowth}
    There exist an observable (Hermitian operator) $\hat{Z}$ such that for a given time $t$ 
    \begin{equation}
        \bra{\alpha}\hat{Z}(t)-\hat{Z}(t_0)\ket{\alpha} = \frac{\epsilon \Delta t}{\gamma},
        \label{eq:thmgamma}
    \end{equation}
    for all $v<1$, where $\gamma$ is the Lorentz factor and $\alpha$ is given by 
    equation (\ref{eq:alphadef}).
\end{theorem}
Observe that the operator $\hat{Z}$ does not depend on $v$: the
dependence on $v$ comes from $\ket{\alpha}$.
\begin{proof}
    Let us take, w.l.o.g. $t_0=0$ and $\hat{Z}(0)=0$. Observe that, given $t$, the function
    \begin{equation}
        \epsilon t\sqrt{1-v^2}e^{|\alpha|^2}=\epsilon t\sqrt{1-v^2}e^{v^2\frac{m}{2\hbar}(t^2\omega+1)}\equiv f(v)
        \label{eq:vfunc}
    \end{equation}
    is holomorphic for all $v<1$. Therefore, there exists a McLaurin series
    \begin{equation}
        \epsilon t\sqrt{1-v^2}e^{v^2\frac{m}{2\hbar}(t^2\omega+1)}= \sum_{k=0}^{\infty}a_kv^{k}.
        \label{eq:SeriesExp}
    \end{equation}
    Observe that for $\hat{Z}$ written in the Harmonic oscillator basis would imply 
    \begin{equation}
        \bra{\alpha}\hat{Z}\ket{\alpha} = e^{-|\alpha|^2}\sum_{n,l}Z_{n,l}\frac{\alpha^{*n}\alpha^l}{\sqrt{n!l!}}.
    \end{equation}
    We can write an expansion of $\bra{\alpha}\hat{Z}\ket{\alpha}e^{|\alpha|^2}$
    dividing it into two terms, 
    \begin{align}
        \sum_{n,l}&Z_{n,l}\frac{\alpha^{*n}\alpha^l}{\sqrt{n!l!}} = \sum_{n=0}^\infty Z_{n,n}\frac{|\alpha|^{2n}}{n!}+\nonumber\\
        &\sum_{n<l}(Z_{n,l}\alpha^{*(n-l)}+Z_{n,l}^*\alpha^{(n-l)})\frac{|\alpha|^{2l}}{\sqrt{n!l!}}, 
        \label{eq:ZHermExp}
    \end{align}
    where $Z_{l,n}=Z_{n,l}^*$ because $\hat{Z}$ is an Hermitian operator. 
    Let us define
    \begin{equation}
        \beta = \sqrt{\frac{m}{2\hbar}}(t\sqrt{\omega}+\frac{i}{\sqrt{\omega}}). 
        \label{eq:betadef}
    \end{equation}
    Therefore Eq. (\ref{eq:ZHermExp}) can be written as 
    \begin{align}
        \sum_{n,l}&Z_{n,l}\frac{\alpha^{*n}\alpha^l}{\sqrt{n!l!}} = \sum_{n=0}^\infty Z_{n,n}\frac{|\beta|^{2n}}{n!}v^{2n}+\nonumber\\
        &\sum_{n<l}2\Re(Z_{n,l}\beta^{*(n-l)})\frac{|\beta|^{2l}}{\sqrt{n!l!}}v^{n+l}. 
    \end{align}
    As we have the freedom to choose the components of $Z_{n,l}$ we do so
    in a manner that they correspond to the $n$th derivative of $f(v)$ (Eq. (\ref{eq:vfunc})), 
    i.e. $f^{(n)}(v=0)$.
    In other words, relabeling we get a series expansion of $\bra{\alpha}\hat{Z}\ket{\alpha}e^{-|\alpha|^2}$
    in terms of $v$, thus condition of Eq. (\ref{eq:SeriesExp}) is fulfilled and
    an operator that behaves as in equation (\ref{eq:thmgamma}) must exist.
\end{proof}
We can illustrate how a QTR machine would work.
Suppose we are given $N_0$ samples
of a source in reference frame $\mathcal{O}$, then the height of the average
would be $N_0(\epsilon\delta)$. 
If samples come from another reference frame where $\gamma> 1$ the steps 
to reach the same height would be given by $N(\epsilon\delta)/\gamma$ for
a $N>N_0$.
As observed in Fig (\ref{fig:clockticks}), the height of the
whole set of measurements yields the record of passed time.
In other words, the number of lines it crosses would be the ticks of
the clock.
Observe that equating the heights in
both cases yield the number of smaller steps $N$ (blue line),
\begin{equation}
    N = \gamma N_0.
\end{equation}
$N_0$ (red line) and $N$  are proportional to measured time, i.e. 
\begin{equation}
    T = \gamma T_0,
\end{equation}
where $T_0$ is time measured in the reference frame $\mathcal{O}$ and $T$ in $\mathcal{O}^\prime$. 
Therefore, to $\mathcal{O}$, the reference frame $\mathcal{O}^\prime$ is
slower. This is all in terms of the information acquisition.

\section{A quantum origin of Lorentz transformations}
\label{sec:AQOrigin}
As we have observed, a Lorentz covariant QTR can be built as a
theoretical machine. However, we propose here to elevate such a mechanism
into a fundamental aspect of matter. 
We need to generalize the QTR family so that we can find more observables
whose expectation values are functions of time which are Lorentz covariant.
To be more precise, from Fig. (\ref{fig:iidsrc}) suppose we want to measure
a quantity $g(t)$ that is time-dependent. If the source is in 
the reference frame $\mathcal{O}$ we would measure $g(t_0)$ at
time $t_0$. We can show that there are observables $\hat{Z}_g$
such that they record a dilated time, i.e. those that
fulfill $g(t/\gamma)=g(t_0)$ if the source is in reference
frame $\mathcal{O}^\prime$ and $g$ holomorphic. We would have that 
$t=t_0\gamma$ for one-to-one functions. 
This is done by generalizing theorem (\ref{thm:linealgrowth}) as follows
\begin{theorem}
    \label{thm:holofunc}
    For any holomorphic function $g(t/\gamma)$ there exist an 
    observable $\hat{Z}_g$ such that for a given time $t$ 
    \begin{equation}
        \bra{\alpha}\hat{Z}_g(t)\ket{\alpha} = g\left(\frac{t}{\gamma}\right),
    \end{equation}
    for all $v<1$. 
\end{theorem}
\begin{proof}
    Observe that, given $t$, the function
    \begin{equation}
        g\left(\frac{t}{\gamma}\right)e^{|\alpha|^2}=g(t\sqrt{1-v^2})e^{v^2\frac{m}{2\hbar}(t^2\omega+1)}
    \end{equation}
    is holomorphic for all $v<1$. Therefore, there exists a McLaurin series
    \begin{equation}
        g(t\sqrt{1-v^2})\sqrt{1-v^2}e^{v^2\frac{m}{2\hbar}(t^2\omega+1)}= \sum_{k=0}^{\infty}a_kv^{k}.
    \end{equation}
    The rest of the proof is analogous to that of theorem \ref{thm:linealgrowth}.
\end{proof}

Now that we know how a QTR machine would work we can state a theory for time dilation.
All matter must have a QTR mechanism: it distinguishes
amongst matter from other reference frames because its information is distorted
and acquiring that information is slower analogous to Fig. (\ref{fig:clockticks}).
We can summarize this theory in the following alternative form of 
postulate \ref{pst:QLrntzPost}:
\begin{postulate}
    All matter has a QTR mechanism which is the source of time dilation amongst
    different reference frames.
    \label{pst:QTRpost}
\end{postulate}
\section{Relativistic momentum and mass}
\label{sec:Cors}
As a corollary of theorem \ref{thm:holofunc} we
can show the existence of the relativistic momentum, Eq. (\ref{eq:RelMom}) and the
relativistic mass, Eq. (\ref{eq:RelMass}). Observe that the momentum operator
expressed in the Harmonic oscillator basis is
\begin{equation}
    \hat{p} = 
i\sqrt{\frac{\hbar m\omega}{2}}\begin{pmatrix}
    0 & -1 & 0 & 0 &\ldots &  \\
    1 & 0 & -\sqrt{2}& 0 & \ldots &  \\
    0 & \sqrt{2}  & 0 & -\sqrt{3} & \ldots & \\
    0 & 0 & \sqrt{3} & 0 & \ldots & \\
    \vdots & & & & \ddots &\\
     & & & & & 
\end{pmatrix}.
\end{equation}
We can write this operator as
\begin{equation}
    \hat{p} = i\sqrt{\frac{\hbar m\omega}{2}} P,
\end{equation}
where $P$ is an infinite matrix whose components are
\begin{equation}
    P_{n,l}=
\begin{cases}
-\sqrt{l}~\text{for}~n-l=-1,\\
\sqrt{l}~\text{for}~l-n=-1\\
0~\text{otherwise}.
\end{cases}
\end{equation}
We have the following corollary of theorem \ref{thm:holofunc}
\begin{corollary}
    \label{cor:RelMom}
There exist an operator $\hat{p}^{Rel}$ such that
\begin{equation}
    \langle\alpha|\hat{p}^{Rel}|\alpha\rangle = \frac{vm}{\sqrt{1-v^2}~},
\end{equation}
for the coherent states with $\alpha$ defined in
Eq. (\ref{eq:alphadef}).
\end{corollary}
\begin{proof}
Following the expansion of the proof of theorem \ref{thm:linealgrowth}  
we define the following infinite matrix $P^{Rel}$ componentwise
\begin{equation}
    P^{Rel}_{n,l}=
\begin{cases}
-\frac{\sqrt{l}f_{l-1}(l-1)!}{|\beta|^{2(l-1)}2(l-1)!}~\text{for}~n-l=-1,\\
\frac{\sqrt{l}f_{l-1}(l-1)!}{|\beta|^{2(l-1)}2(l-1)!}~\text{for}~l-n=-1\\
0~\text{otherwise},
\end{cases}
\end{equation}
where $\beta$ is defined in Eq. (\ref{eq:betadef}) and $f_l$ are defined as
\begin{equation}
    f_l \equiv \left.\frac{d^{2l}}{dv^{2l}}\left(\frac{e^{v^2|\beta|^2}}{\sqrt{1-v^2}}\right)\right|_{v=0}.
\end{equation}
Let us define
\begin{equation}
    \hat{p}^{Rel} \equiv i\sqrt{\frac{\hbar m\omega}{2}} P^{Rel}.
\end{equation}
We have that, as there are no diagonal terms
    \begin{align}
     &\langle\alpha|\hat{p}^{Rel}|\alpha\rangle\nonumber\\ 
&= e^{-|\alpha|^2}\sqrt{\frac{\hbar m\omega}{2}}\sum_{l=1}^\infty2\Re\left(\frac{-i\sqrt{l}f_{l-1}(l-1)!\beta}{|\beta|^{2(l-1)}2(l-1)!|\beta|^2}\right)\times\\
&\times\frac{|\beta|^{2l}v^{2(l-1)}v}{\sqrt{(l-1)!l!}},\\
&= e^{-|\alpha|^2}vm\sum_{l=1}^\infty\frac{f_{l-1}v^{2(l-1)}}{2(l-1)!}\nonumber,\\
        &= e^{-|\alpha|^2}vm  \frac{e^{|\alpha|^2}}{\sqrt{1-v^2}}\nonumber.
    \end{align}
\end{proof}
Likewise, we can prove the existence of a relativistic mass operator 
$\hat{\mu}^{Rel}$. 
\begin{corollary}
    \label{cor:RelMass}
There exist an operator $\hat{\mu}^{Rel}$ such that
\begin{equation}
    \langle\alpha|\hat{\mu}^{Rel}|\alpha\rangle = \frac{m}{\sqrt{1-v^2}~},
\end{equation}
for the coherent states with $\alpha$ defined in
Eq. (\ref{eq:alphadef}).
\end{corollary}
\begin{proof}
    Let us define the diagonal operator
    \begin{equation}
        \mu^{Rel}_{n,n}=m\frac{n!f_n}{2n!|\beta|^{2n}}.
    \end{equation}
Observe that following the expansion made in theorem \ref{thm:linealgrowth}
we get
    \begin{align}
     &\langle\alpha|\hat{\mu}^{Rel}|\alpha\rangle\nonumber\\ 
        &= e^{-|\alpha|^2}\sum_{n=0}^\infty m\frac{n!f_n}{2n!|\beta|^{2n}}\frac{|\beta|^{2n}}{n!}v^{2n} ,\\
        &= e^{-|\alpha|^2}\sum_{n=0}^\infty m\frac{f_n}{2n!}v^{2n} \nonumber,\\
        &= e^{-|\alpha|^2}m\frac{e^{|\alpha|^2}}{\sqrt{1-v^2}} \nonumber.\\
    \end{align}
\end{proof}
\section{Transparency and the twin paradox}
\label{sec:TransAndTwin}
To elevate these investigations into a fundamental perspective
of nature, several clarifications are in order.
Observe that this is not a theory that eliminates the concept of spacetime, but rather 
allocates it within a quantum formalism without imposing it as a constriction 
as it is usually done. As such it does not eliminate the locality of interactions.  
However, this theory does not imply a restriction in velocities as Special
Relativity does. To recover this important constraint we impose a 
\emph{transparency} constraint: any faster-than-light
particles would be transparent to any detector. 
This constraint is already present in Fig. (\ref{fig:avsgrow})
as a particle in a reference frame at light speed with respect to the measuring one
would cause no information. 

The phenomenon of transparency has been common in
quantum physics since almost its inception. Fermi observed the 
necessity of slowing
down neutrons for them to interact with heavy nuclei, which served to
induce radioactivity and lead to nuclear engineering \cite{Fermi1934Artificial}.
In other words, the heavy nuclei are transparent to the passage
of fast neutrons. Also, the neutrino is an elusive particle that is 
hard to detect because matter is almost transparent to them. There
are needed enormous machines to detect them like the Super-Kamiokande
which detected neutrino oscillations \cite{EvidenceForOsFukuda1998}.

A second relevant issue is that we are making a description in terms of what
is usually called the coordinate time. 
As it is well known in Relativity
literature, there is a difference concerning the physical and detectable
proper time \cite{schutz_2009}. For us, the proper time would be the one
measured by the detector, from Fig. (\ref{fig:iidsrc}) the one in reference
frame $\mathcal{O}$.
The difference between coordinate and proper time becomes clear in the known 
``twin paradox'': twin A stays on planet Earth while twin B takes a spaceship
at $v=0.96$ (remember that in our units $v=1$ is the speed of light) for seven years.
Then, the twin B turns around instantly and returns to planet Earth at the same speed.
When returning home twin B has aged 14 years while twin A has aged 50 years.
A link with our approach can be made by observing that, in the go trajectory, 
from B's perspective
all the quantum state of the planet Earth would be distorted and if B could measure something
from Earth's reference frame it would follow time dilation according to
Lorentz transformations. When twin B's ship changes its reference frame
to return to Earth, then, because of simultaneity, twin A would age in an instant.
While returning home the distortion would correspond to the new reference frame.
If twin B could measure the quantum state of planet Earth, it would be distorted
as well following time dilation. The path traversed in spacetime would be the
same, since we are not eliminating spacetime but merely giving a quantum origin 
for it.
\section{Discussion}
\label{sec:disc}
We have thus shown the existence of QTRs where time dilates according to 
Lorentz transformations. 
We have done it for the special case of the Harmonic Oscillator.
The ground state of a Harmonic Oscillator looks like a nontrivial coherent 
state in a different reference frame that moves at constant speed $v$ with respect to
the one where it is measured. Time dilation would thus be a phenomenon caused by the
distortion of states in different reference frames for Lorentz-covariant QTRs.  

Observe that time dilation is something usually associated with spacetime as a
field, outside matter itself. 
Here, we make the statement that all matter
has a QTR mechanism which is the origin of time dilation expressed in
postulate \ref{pst:QTRpost}. Thus, we rebuild Special Relativity from a
quantum origin.

This all originates from a fundamental change of perspective: we consider the acquisition
of information as the source of time, or more succinctly, change produces time.
In other approaches, time is an intrinsic property 
of the universe that would cause the change: exactly the opposite.
Time produces change is the perspective of Pauli or 
the relational physics approach \cite{PauliSTheoremGalapo2002,QuantumTimeGiovan2015}.

We introduced an operator for relativistic momentum $\hat{p}^{Rel}$ 
and mass $\hat{\mu}^{Rel}$. These operators suggest that an equation
similar to Dirac's can be found.

Perhaps the most radical outcome of this proposal is the transparency
constraint that comes from allowing the existence of faster-than-light
massive particles but with the condition that they become unobservable
by the QTRs in $\mathcal{O}$, i.e. matter from $\mathcal{O}$ becomes
transparent to them. The phenomenon of transparency is however common in quantum physics, as
Fermi observed that neutrons needed to be slowed down because otherwise
heavy nuclei would be transparent to them. Also the ellusive neutrinos
hardly interact with matter, which makes matter almost transparent to 
them.  

We are proposing here a deviation from the usual spacetime picture.
We suggest that the mixture happens in the interaction: 
time measured from a source that comes from another reference frame would become distorted.
This has an advantage in the sense that it provides a clear explanation of spacetime
from the perspective of Quantum Mechanics. However, the long-range aim of 
these inquiries is
to provide a coherent theory of Quantum Gravitation. 
We do not make use of a field in this approach to Special Relativity.
However, there are important gravitational 
phenomena that can only be understood with a field, like gravitational waves \cite{ObservationOfAbbott2016}.
There must be a field of some sort so that information can be transmitted
through it. It does not conflict with a theory that states that spacetime effects 
are originated
from a QTR mechanism. Inclusion of field effects in matter can be made
considering QTRs that depend on an interaction with a field i.e. $\hat{Z}_g(\kappa)$ where
$\kappa$ is a parameter of a field. Perhaps an 
extension of the transparency effect can help to
understand dark matter: the gravitational effects are present but the matter does
not interact with anything. However, a General theory is needed. This is a matter of future study.
\bibliography{bibliography}
\end{document}